\documentclass[conference,letterpaper]{IEEEtran}
\usepackage[utf8]{inputenc} 
\usepackage[T1]{fontenc}
\usepackage{url}
\usepackage{ifthen}
\usepackage{cite}
\usepackage[cmex10]{amsmath} 
\usepackage{mathpple}
\usepackage{times}
\usepackage{amsthm,xpatch}
\usepackage{amsfonts}
\usepackage{amssymb}
\usepackage{breqn}
\usepackage{mathtools}
\usepackage{bbm}
\usepackage{comment}
\usepackage{multirow}

\newtheorem{theorem}{Theorem}

\newtheorem{remark}{Remark}

\newtheorem*{example*}{Example}

\renewcommand{\baselinestretch}{0.98}

\IEEEoverridecommandlockouts

\begin{document}

\makeatletter
\renewenvironment{proof}[1][\proofname]{\par
  \pushQED{\qed}%
  \normalfont \topsep6\p@\@plus6\p@\relax
  \trivlist
  \item[\hskip\labelsep
        \itshape
    #1\@addpunct{:}]\ignorespaces
}{
  \popQED\endtrivlist\@endpefalse
}
\makeatother

\pagestyle{plain}

\title{A Linear Programming Approach to\\ Private Information Retrieval}

\author{%
  \IEEEauthorblockN{Anoosheh Heidarzadeh}
  \IEEEauthorblockA{Department of Electrical and Computer Engineering \\
                    Santa Clara University\\
                    Santa Clara, CA 95053 USA\\
                    Email: aheidarzadeh@scu.edu}
  \and
  \IEEEauthorblockN{Ningze Wang and Alex Sprintson}
  \IEEEauthorblockA{Department of Electrical and Computer Engineering\\
                    Texas A\&M University\\
                    College Station, TX 77843 USA\\
                    Email: \{ningzewang, spalex\}@tamu.edu}
}

\maketitle 

\thispagestyle{plain}

\begin{abstract}
This work presents an algorithmic framework that uses linear programming to construct \emph{addition-based Private Information Retrieval (AB-PIR)} schemes, where retrieval is performed by downloading only linear combinations of message symbols with coefficients set to 0 or 1. The AB-PIR schemes generalize several existing capacity-achieving PIR schemes and are of practical interest because they use only addition operations---avoiding multiplication and other complex operations---and are compatible with any finite field, including binary. Our framework broadens the search space to include all feasible solutions and can be used to construct optimal AB-PIR schemes for the entire range of problem parameters, including the number of servers, the total number of messages, and the number of messages that need to be retrieved. The framework enables us to identify schemes that outperform the previously proposed PIR schemes in certain cases and, in other cases, achieve performance on par with the best-known AB-PIR solutions. Additionally, the schemes generated by our framework can be integrated into existing solutions for several related PIR scenarios, improving their overall performance.
\end{abstract}

\section{Introduction}
In this work, we revisit the problem of \emph{Private Information Retrieval (PIR)}, where a user is interested in retrieving one or more messages from a dataset stored on remote servers. 
The user's goal is to download the minimum amount of information required while revealing no information about the identities of the desired messages to any of the servers. 

Several variations of PIR have been studied in recent years. 
This includes PIR with non-colluding servers (see, e.g.,~\cite{SJ2017,SJ2016ArbitraryTIFS,SJ2018Multiround,ZX2019,TSC2019,BU2018,WHS2022,VBU2022}), PIR with colluding servers (see, e.g.,~\cite{SJ2018Colluding,BU2019Colluding,LJJ2021,HFLH2022}), and PIR with coded databases (see, e.g.,~\cite{TGKHHER2017,BU18,BAWU2020}). 
Among these, two notable variants are single-message PIR~\cite{SJ2017} and 
multi-message PIR~\cite{BU2018}. 
In both settings, $N$ non-colluding servers store identical copies of $K$ messages, and a user wants to privately retrieve $D$ out of the $K$ messages, where the case of $D=1$ corresponds to the single-message setting and the case of $D\geq 2$ corresponds to the multi-message setting. 
The objective in both settings is to maximize the rate, defined as the ratio of the number of bits required by the user to the total number of bits downloaded from all servers. The maximum achievable rate in each setting is referred to as its capacity.

In~\cite{SJ2017}, Sun and Jafar presented capacity-achieving schemes for single-message PIR for all values of $N$ and $K$. 
Building on these, Banawan and Ulukus in~\cite{BU2018} proposed PIR schemes for the multi-message setting. 
While these schemes are optimal for certain values of $N$, $K$, and $D$, their optimality for other values of $N$, $K$, and $D$ remained unknown. 

In this work, we focus on a class of PIR schemes, referred to as \emph{addition-based PIR (AB-PIR)} schemes, where the user downloads only linear combinations of message symbols from the servers, with combination coefficients set to 0 or 1.
Several existing capacity-achieving PIR schemes belong to this class, including the scheme in~\cite{SJ2017} for $D=1$ and the scheme in~\cite{BU2018} for $D\leq \frac{K}{2}$. 
AB-PIR schemes are particularly interesting for practical use, as they operate over any finite field, including binary, use only addition operations and avoid multiplication or other complex operations, and reduce upload cost by eliminating the need to communicate combination coefficients. 

The main contribution of the paper is an algorithmic framework based on linear programming (LP) that can be used to identify optimal AB-PIR schemes for all values of $N$, $K$, and $D$.
Specifically, we introduce a general class of AB-PIR schemes characterized by a set of parameters, which can be optimized via an LP formulation to identify an AB-PIR scheme that achieves the maximum rate.
When ${D\mid K}$, our framework yields an optimal scheme, similar to the AB-PIR scheme in~\cite{BU2018}. 
More interestingly, when ${D\nmid K}$, for certain values of $N$, $K$, and $D$, it yields schemes that outperform the one in~\cite{BU2018}, while for other values, they perform similarly to the one in~\cite{BU2018}. 
For instance, when $K=5$ and $D=2$, the schemes generated by our framework achieve rates of $\frac{82}{135}\approx 0.6074$ and $\frac{57}{80}= 0.7125$ for $N=2$ and $N=3$, respectively. 
In contrast, the best previously-known achievable rates for the same problem parameters, due to the scheme in~\cite{BU2018}, were $\frac{17}{28}\approx 0.6071$ and $\frac{42}{59}\approx 0.7119$, respectively. 

Additionally, our schemes can substitute the one in~\cite{BU2018}, which serves as a building block in existing solutions for several related scenarios, such as multi-message PIR with private side information~\cite{SSM2018} and private inner-product retrieval~\cite{MMM2019}, yielding more efficient solutions for these scenarios.

While our framework is applied to AB-PIR schemes in this work, it can be extended to more general classes of PIR schemes that use non-binary coefficients and may yield schemes that outperform the best-known solutions in other PIR settings.
Although not discussed here, the framework can also be used to establish converse bounds and construct achievability schemes under structural constraints, such as linearity or limited subpacketization, in various PIR scenarios.

\section{Problem Setup}\label{sec:SN}
We represent random variables and their realizations by bold-face symbols and regular symbols, respectively. 
For an integer $i\geq 1$, 
the set $\{1,\dots,i\}$ is denoted by $[i]$, and for integers $1\leq i<j$, 
the set $\{i,\dots,j\}$ is denoted by $[i:j]$. 

Consider $N$ non-colluding servers, each storing an identical copy of $K$ messages ${\mathrm{X}_1,\dots,\mathrm{X}_K}\in \mathbbmss{F}^m_q$. 
The random variables $\mathbf{X}_1,\dots,\mathbf{X}_K$ are assumed to be independent and uniformly distributed over $\mathbbmss{F}^{m}_{q}$. 
That is, each message $\mathrm{X}_i$ consists of $m$ symbols from a finite field $\mathbbmss{F}_q$. 
We refer to $m$ as the message length and $q$ as the field size. 

Consider a user who wants to retrieve $D$ (out of $K$) messages, denoted as $\mathrm{X}_{\mathrm{W}}:=\{\mathrm{X}_i: i\in \mathrm{W}\}$, where $\mathrm{W}$ is a $D$-subset of $[K]$. 
The random variable $\mathbf{W}$ is uniformly distributed over all $D$-subsets of ${[K]}$ and is independent of $\mathbf{X}_1,\dots,\mathbf{X}_K$. 
We refer to $\mathrm{X}_{\mathrm{W}}$ as the \emph{demand messages} and 
$\{\mathrm{X}_i: i\in [K]\setminus \mathrm{W}\}$ as the \emph{interference messages}.  

For each ${n\in [N]}$, the user generates a query $\mathrm{Q}^{[\mathrm{W}]}_n$, and sends it to server $n$. 
Each query $\mathrm{Q}^{[\mathrm{W}]}_n$ is a (deterministic or stochastic) function of the demand's index set $\mathrm{W}$ and is independent of the messages $\mathrm{X}_{[K]} := \{\mathrm{X}_i: i\in [K]\}$. 
Moreover, each query $\mathrm{Q}^{[\mathrm{W}]}_n$ must reveal no information about the demand's index set $\mathrm{W}$ to server $n$, i.e., 
\begin{align}\label{eq:PC}
I(\mathbf{W}; \mathbf{Q}^{[\mathbf{W}]}_n) = 0, \quad \forall n\in [N].
\end{align}
 
Subsequently, each server $n$ generates an answer $\mathrm{A}^{[\mathrm{W}]}_n$ and sends it to the user. 
Each answer $\mathrm{A}^{[\mathrm{W}]}_n$ is a deterministic function of the query $\mathrm{Q}^{[\mathrm{W}]}_n$ and the messages $\mathrm{X}_{[K]}$. 
Moreover, the user must be able to recover the demand messages $\mathrm{X}_{\mathrm{W}}$ given the collection of answers
${\mathrm{A}^{[\mathrm{W}]}_{[N]}:=\{\mathrm{A}^{[\mathrm{W}]}_n:n\in [N]\}}$ and queries ${\mathrm{Q}^{[\mathrm{W}]}_{[N]}:=\{\mathrm{Q}^{[\mathrm{W}]}_n: n\in [N]\}}$, 
i.e., 
\begin{equation}\label{eq:RC}
H(\mathbf{X}_{\mathrm{W}}| \mathbf{A}^{[\mathrm{W}]}_{[N]},\mathbf{Q}^{[\mathrm{W}]}_{[N]})=0.
\end{equation}

The problem is to design a scheme for generating the queries ${\mathrm{Q}^{[\mathrm{W}]}_{[N]}}$ and the corresponding answers ${\mathrm{A}^{[\mathrm{W}]}_{[N]}}$ for any given $\mathrm{W}$ such that both the privacy and recoverability conditions defined in~\eqref{eq:PC} and~\eqref{eq:RC} are satisfied. 
This problem, initially studied in~\cite{SJ2017} for the case of $D=1$ and later extended in~\cite{BU2018} for the cases of $D\geq 2$, is known as \emph{single-message PIR} when $D=1$ and \emph{multi-message PIR} when ${D\geq 2}$. 
We refer to both scenarios collectively as PIR. 

In this work, we focus on \emph{addition-based PIR (AB-PIR) schemes}, where each server's answer to the user's query consists only of linear combinations of message symbols with coefficients restricted to $0$ and $1$. 
We define the \emph{rate} of an AB-PIR scheme as the ratio of the number of bits required by the user, $H(\mathbf{X}_{\mathrm{W}})$, 
to the total number of bits downloaded from all servers, ${\sum_{n\in [N]} H(\mathbf{A}^{[\mathrm{W}]}_n|\mathbf{Q}^{[\mathrm{W}]}_n)}$. 
Additionally, we define the \emph{capacity} of the AB-PIR problem as the maximum achievable rate among all AB-PIR schemes and refer to a capacity-achieving scheme as \emph{optimal}. 

Our goal is to design optimal AB-PIR schemes that are applicable for all values of $N$, $K$, and $D$, and  work with any field size $q$, as the message length $m$ grows large.  

\section{Main Results}
In this section, we present our main results on the capacity of the AB-PIR problem for all $N$, $K$, and $D$.  

To simplify the notation, 
let $\mathrm{v}_1,\dots,\mathrm{v}_K$ be $K$ vectors, each of length $D$, where for each ${s\in [1:K-D]}$, ${\mathrm{v}_s = \frac{1}{N-1}\sum_{t=1}^{D}\binom{D}{t} \mathrm{v}_{t+s}}$, 
and for each ${s\in [K-D+1:K]}$, $\mathrm{v}_{s}$ is a unit vector whose $(s-K+D)$th component is $1$ and all other components are $0$. 
Moreover, let $\mathrm{f}$ and $\mathrm{g}$ be two vectors, each of length $D$, defined as
\begin{equation}\label{eq:f}
\mathrm{f} := \frac{N}{D}\sum_{s=1}^{K} \binom{K}{s}\mathrm{v}_s, 
\end{equation}
and
\begin{equation}\label{eq:g}
\mathrm{g} := \mathrm{f} - \frac{N}{D}\sum_{s=1}^{K-D}\binom{K-D}{s}\mathrm{v}_s.
\end{equation}

\begin{theorem}\label{thm:1}
The capacity of AB-PIR with $N$ servers, $K$ messages, and $D$ demand messages is lower bounded by
\begin{equation}\label{eq:RLB}
\underline{R}:=\max\left\{\frac{g_1}{f_1},\dots,\frac{g_D}{f_D}\right\},
\end{equation} 
where $\mathrm{f} = [f_1,\dots,f_D]$ and $\mathrm{g} = [g_1,\dots,g_D]$ are defined in~\eqref{eq:f} and~\eqref{eq:g}, respectively, 
and is upper bounded by \begin{equation}\label{eq:RUB2}
\overline{R}:=\left(\frac{1-1/N^{\lfloor {K}/{D}\rfloor}}{1-1/N}+\frac{K/D-\lfloor K/D \rfloor}{N^{\lfloor{K}/{D}\rfloor}}\right)^{-1}.    
\end{equation}
\end{theorem}

The upper bound in~\eqref{eq:RUB2}, which appears without proof, follows directly from the converse results in~\cite{BU2018} for all PIR schemes. 
To establish the lower bound in~\eqref{eq:RLB}, we propose a new AB-PIR scheme that achieves this rate. 
First, we introduce a general class of AB-PIR schemes, which are characterized by a set of parameters. 
We then optimize these parameters using linear programming to maximize the rate. 

\begin{theorem}\label{thm:2}
The proposed AB-PIR scheme outperforms the one in~\cite{BU2018} when ${g_t/f_t > g_D/f_D}$ for some ${t\in [D-1]}$ and performs similarly when ${g_t/f_t\leq g_D/f_D}$ for all ${t\in [D-1]}$. 
\end{theorem}

The proof relies on the fact that the scheme in~\cite{BU2018} belongs to the class of AB-PIR schemes over which we perform optimization to identify one with the maximum rate.  

\begin{remark}\label{rem:1}\normalfont
The AB-PIR scheme in~\cite{BU2018} was previously shown to achieve capacity when $D\mid K$. 
This directly implies that our scheme is also optimal in these cases, i.e., the lower bound $\underline{R}$ in~\eqref{eq:RLB} and the upper bound $\overline{R}$ in~\eqref{eq:RUB2} are equal when $D\mid K$. 
In contrast, when $D\nmid K$, we observe that $\underline{R}<\overline{R}$. 
Numerical results show that $\underline{R}\geq 0.9868\overline{R}$ for $D<\frac{K}{2}$ and $\underline{R}\geq 0.9621\overline{R}$ for $D>\frac{K}{2}$. 
Nevertheless, for $D\nmid K$, a theoretical characterization of the gap between $\underline{R}$ and $\overline{R}$ is unavailable, and it is unclear whether the upper bound, the lower bound, or both are loose for AB-PIR schemes. 
\end{remark}

\begin{remark}\label{rem:2}\normalfont
Our analysis, presented in the appendix, shows that there are infinitely many problem instances where our scheme outperforms the one in~\cite{BU2018}.
Specifically, for ${D=2}$, our scheme is superior for all odd $K$ and any $N$.
However, fully characterizing the instances where our scheme is superior for $D\geq 3$ remains an open problem. 
\end{remark}

\section{Proofs of Theorems}\label{sec:SCHEME}
\subsection{Proof of Theorem~\ref{thm:1}}
We prove the achievability part of Theorem~\ref{thm:1} by presenting an AB-PIR scheme that achieves the rate $\underline{R}$ defined in~\eqref{eq:RLB}. 
The proposed scheme applies to all values of $N$, $K$, and $D$, works with any field size $q$, and requires a sufficiently large message length $m$, determined by solving an LP problem. 

The proposed scheme operates on message subpackets, with each message divided into $L$ subpackets, each containing $m/L$ symbols from the $m$ symbols of the message, where the choice of $L$ will be determined later. 
Specifically, the answer of each server includes $L_1$ subpackets from each message and, for each $s\in [2:K]$, $L_s$ sums of subpackets from every $s$-subset of messages, with  
each subpacket either not contributing to the server's answer or appearing exactly once. 
The choice of $L_1,\dots,L_K$ will be determined later.  

More specifically, each server's answer includes $L_1$ singletons for each message, where each singleton is a distinct subpacket of the message. 
Additionally, for each ${i\in [2:D]}$, the answer includes $L_i$ $(i,0)$-sums for every $i$-subset ${\mathcal{I}\subseteq \mathrm{W}}$, where each $(i,0)$-sum consists of $i$ subpackets, each from a distinct demand message in $\mathrm{X}_{\mathcal{I}}$. 
Similarly, for each ${j\in [2:K-D]}$, the answer includes $L_{j}$ $(0,j)$-sums for every $j$-subset ${\mathcal{J}\subseteq [K]\setminus \mathrm{W}}$, where each $(0,j)$-sum consists of $j$ subpackets, each from a distinct interference message in $\mathrm{X}_{\mathcal{J}}$. 
Moreover, for each ${i\in [D]}$ and ${j\in [K-D]}$, the answer includes $L_{i+j}$ $(i,j)$-sums for every $i$-subset $\mathcal{I}\subseteq \mathrm{W}$ and $j$-subset $\mathcal{J}\subseteq [K]\setminus \mathrm{W}$, where each $(i,j)$-sum comprises $i$ subpackets from the demand messages in $\mathrm{X}_{\mathcal{I}}$ and $j$ subpackets from the interference message in $\mathrm{X}_{\mathcal{J}}$. 

Since the scheme requires downloading a total of ${M:=\sum_{s=1}^{K} \binom{K}{s} L_s}$ singletons and sums from each server, 
the number of downloads per demand message is given by 
\begin{dmath}\label{eq:DC}
\frac{NM}{DL} = \frac{N}{DL}\sum_{s=1}^{K}\binom{K}{s} L_s. \end{dmath}
To maximize the rate, we need to minimize the number of downloads per demand message in~\eqref{eq:DC}. 
Our goal is then to solve this optimization problem with respect to the variables $L_1,\dots,L_K$ and $L$, while ensuring privacy and recoverability. 

Since every message---whether a demand message or an interference message---contributes equally to each server's answer, 
the privacy condition is inherently satisfied and does not impose any constraints on the values of $L_1,\dots,L_K$, or $L$. 
However, as we will discuss shortly, $L_1,\dots,L_K$, and $L$ must satisfy certain constraints to ensure that each demand message can be fully recovered from the servers' answers. 

Since there exist $\binom{K-D}{j}$ $j$-subsets of interference messages for each $j\in [K-D]$, the user can cancel the interference part of each $(i,j)$-sum which is aligned with either a singleton pertaining to an interference message (i.e., ${j=1}$) or a $(0,j)$-sum pertaining to $j$ interference messages (i.e., ${j\in [2:K-D]}$) retrieved from another server. 
This results in the recovery of $\binom{K-D}{j} L_{1+j}$ new singletons for every demand message and, for each $i\in [2:D]$, $\binom{K-D}{j} L_{i+j}$ new $(i,0)$-sums for every $i$-subset of demand messages. 

We note that this is subject to the condition that the total number of $(i,1)$-sums (or $(i,j)$-sums for each ${j\in [2:K-D]}$) for all $i\in [D]$, retrieved from a server and corresponding to the same interference message (or $j$-subset of interference messages), must not exceed the number of singletons corresponding to that interference message (or $(0,j)$-sums corresponding to those $j$ interference messages), retrieved from all other servers.  
Thus, it must hold that 
\begin{equation}\label{eq:C0}
L_j \geq \frac{1}{N-1}\sum_{i=1}^D \binom{D}{i} L_{i+j}, \quad \forall j\in [K-D].
\end{equation} 
While this is an inequality constraint, we impose it as an equality constraint to simplify the analysis, i.e.,  
\begin{equation}\label{eq:C1}
L_j = \frac{1}{N-1}\sum_{i=1}^D \binom{D}{i} L_{i+j}, \quad \forall j\in [K-D].
\end{equation} 
Our numerical studies suggest that any optimal solution satisfying the constraint in~\eqref{eq:C0} also satisfies the constraint in~\eqref{eq:C1}, but a formal proof has yet to be established.

In summary, from each server's answer, the user can recover a total of $\sum_{j=0}^{K-D} \binom{K-D}{j} L_{1+j}$ singletons for every demand message and, for each ${i\in [2:D]}$, $\sum_{j=0}^{K-D} \binom{K-D}{j} L_{i+j}$ $(i,0)$-sums for every $i$-subset of demand messages, 
provided the condition in~\eqref{eq:C1} is satisfied. 

Singletons corresponding to demand messages---retrieved from each server---can directly be used to recover
\begin{equation}\label{eq:R1}
{R_1 := \sum_{j=0}^{K-D} \binom{K-D}{j} L_{1+j}}
\end{equation}
subpackets of every demand message from that server's answer. 
Additionally, by using a $(2,0)$-sum, which corresponds to a given pair of demand messages and is recovered from a server's answer, along with a subpacket of either message in the pair---retrieved from another server, 
the user can recover a new subpacket of the other message in the pair.
Since every demand message appears in $\binom{D-1}{1}\sum_{j=0}^{K-D} \binom{K-D}{j} L_{2+j}$ $(2,0)$-sums recovered from each server's answer, 
the user can recover 
${\frac{1}{2} \binom{D-1}{1}\sum_{j=0}^{K-D} \binom{K-D}{j} L_{2+j}}$ new subpackets of every demand message from each server's answer, in addition to the $R_1$ subpackets retrieved as singletons. 

In general, for each ${i\in [2:D]}$, using $(i,0)$-sums recovered from a server's answer and the subpackets recovered from other servers' answers, 
the user can recover 
\begin{eqnarray}\label{eq:Mi}
R_i &:=&\frac{1}{i}\binom{D-1}{i-1}\sum_{j=0}^{K-D} \binom{K-D}{j} L_{i+j},
\end{eqnarray}
new subpackets of every demand message. 

Note that $R_1$ in~\eqref{eq:R1} is an integer for any choice of $L_1,\dots,L_K$; however, $L_1,\dots,L_K$ must be selected carefully to ensure that $R_i$ in~\eqref{eq:Mi} is an integer for all $i\in [2:D]$. 

Since the user can recover $\sum_{i=1}^{D} R_i$ distinct subpackets of every demand message from each server, 
they can fully recover each demand message if ${N\sum_{i=1}^{D} R_i = L}$, or equivalently,  
\begin{equation}\label{eq:C3}
\frac{N}{D}\sum_{s=1}^{K}\binom{K}{s}L_s - \frac{N}{D}\sum_{s=1}^{K-D} \binom{K-D}{s} L_{s} = L. 
\end{equation}
 
Our goal is thus to minimize the number of downloads per demand message, as defined in~\eqref{eq:DC}, 
with respect to the variables $L_1,\dots,L_K$, and $L$, while satisfying the constraints specified 
in~\eqref{eq:C1} and~\eqref{eq:C3}. 
Since the objective function is linear in variables $L_1/L,\dots,L_K/L$ and the constraints are linear in variables $L_1,\dots,L_K$, and $L$, 
this optimization problem can be reformulated as an equivalent LP problem by dividing both sides of each constraint by $L$ and replacing $L_s/L$ with $x_s$ for each $s\in [K]$. 
The resulting LP problem can then be solved with respect to the variables $x_1,\dots,x_K\geq 0$:  
\begin{equation*}
\begin{array}{ll@{}ll}
\text{minimize} & \displaystyle \frac{N}{D}\sum_{s=1}^{K} \binom{K}{s} x_s & \\[0.25cm]
\text{subject to} & \displaystyle x_j = \frac{1}{N-1}\sum_{i=1}^{D} \binom{D}{i} x_{i+j}, \quad \forall j\in [K-D] & \\[0.25cm]
& \displaystyle \frac{N}{D}\sum_{s=1}^{K}\binom{K}{s}x_s - \frac{N}{D}\sum_{s=1}^{K-D} \binom{K-D}{s} x_{s} = 1. & 
\end{array}
\end{equation*}

Since the values of $x_1,\dots,x_K$ are uniquely determined given the values of $x_{K-D+1},\dots,x_K$, we can further simplify the LP problem. 
To simplify the notation, we define a vector $\mathrm{v}_s := [v_{s,1},\dots,v_{s,D}]$ for each $s\in [K]$, 
where $v_{s,t}$ for $t\in [D]$ are such that
${x_{s} = \sum_{t=1}^{D} v_{s,t} x_{K-D+t}}$. 
We note that $\mathrm{v}_{K-D+1},\dots,\mathrm{v}_{K}$ are unit vectors, i.e., 
for each ${s\in [K-D+1:K]}$, 
${v_{s,s-K+D} = 1}$ and ${v_{s,t} = 0}$ for all ${t\in [D]\setminus \{s-K+D\}}$. 
Additionally, the vectors $\mathrm{v}_1,\dots,\mathrm{v}_{K-D}$ are uniquely determined given the unit vectors $\mathrm{v}_{K-D+1},\dots,\mathrm{v}_{K}$, 
i.e., for each $s\in [K-D]$, 
${v_{s,t} = \frac{1}{N-1}\sum_{i=1}^{D} \binom{D}{i} v_{i+s,t}}$ for all $t\in [D]$.

Using the vectors  $\mathrm{v}_1,\dots,\mathrm{v}_K$, we can rewrite the LP problem in terms of the variables $x_{K-D+1},\dots,x_{K}\geq 0$: 
\begin{equation*}
\begin{array}{ll@{}ll}
\text{minimize} & \displaystyle \sum_{t=1}^{D} f_t x_{K-D+t} & \\[0.25cm]
\text{subject to} & \displaystyle \sum_{t=1}^{D} g_t x_{K-D+t} = 1. & 
\end{array}
\end{equation*}
where $\mathrm{f} = [f_1,\dots,f_D]$ and $\mathrm{g} = [g_1,\dots,g_D]$ are defined as in~\eqref{eq:f} and~\eqref{eq:g}, respectively. 

An optimal solution to this LP problem is given by ${x_{K-D+t^{*}} = 1/g_{t^{*}}}$, and $x_{K-D+t} = 0$ for all ${t\in [D]\setminus \{t^{*}\}}$, 
where ${t^{*} \in [D]}$ is such that $g_{t^{*}}/f_{t^{*}} = \max_{t\in [D]} g_t/f_t$, and the optimal value of the objective function in this LP problem, $\sum_{t=1}^{D} f_t x_{K-D+t}$, is given by $f_{t^{*}}/g_{t^{*}}$.  
Using $x_{K-D+1},\dots,x_{K}$, 
we can then find $x_1,\dots,x_{K-D}$ as $x_s = \sum_{t=1}^{D} v_{s,t} x_{K-D+t} = v_{s,t^{*}} x_{K-D+t^{*}} = v_{s,t^{*}}/g_{t^{*}}$ for each ${s\in [K-D]}$.  
This yields an optimal solution to the original LP problem in terms of $x_{1},\dots,x_K$, and hence, the optimal value of the objective function, $\frac{N}{D}\sum_{s=1}^{K}\binom{K}{s} x_s$, is given by $f_{t^{*}}/g_{t^{*}}$. 

Since $v_{s,t}$ for all $s\in [K]$ and $t\in [D]$, and consequently, $f_t$ and $g_t$ for all $t\in [D]$, are rational, 
$x_{1},\dots,x_K$ are also rational. 
Using $x_{1},\dots,x_K$, we can thus determine an optimal integral solution $L_1,\dots,L_K$ and $L$ for the original problem by setting $L_s = L x_s$ for all $s\in [K]$ and selecting   
$L$ as a positive integer such that 
(i) 
$L_s$ is an integer for all $s\in [K]$ and (ii) 
$R_i$, as defined in~\eqref{eq:Mi}, is an integer for all ${i\in [2:D]}$. 
Specifically, there exists an integer $1\leq S\leq D$ such that ${L=S(N-1)^{K-D}g_{t^{*}}}$ satisfies both conditions (i) and (ii), since $v_{s,t}$ share ${(N-1)^{K-D}}$ as a common denominator.
Thus, the optimal value of the objective function in~\eqref{eq:DC} is given by $f_{t^{*}}/g_{t^{*}}$, and the rate of the scheme is given by $g_{t^{*}}/f_{t^{*}}$, 
which matches the lower bound $\underline{R}$ defined in~\eqref{eq:RLB}.

\subsection{Proof of Theorem~\ref{thm:2}}
To prove Theorem~\ref{thm:2}, we observe that each variable $L_s$ for $s\in [K]$ in the proposed scheme corresponds to the number of stages used in round $s$ of the scheme in~\cite{BU2018}. 
However, in~\cite{BU2018}, the values of $L_1,\dots,L_K$, and $L$ are not determined through optimization. 
Instead, each $L_j$ for $j\in [K-D]$ is computed using a backward recurrence relation, which coincides with the constraint in~\eqref{eq:C1}, 
with initial conditions $L_{K-D+t} = 0$ for all $t\in [D-1]$ and $L_K = (N-1)^{K-D}$, and the value of $L$ is then determined by an equation that matches the constraint in~\eqref{eq:C3}. 
While these values of $L_1,\dots,L_K$, and $L$ satisfy the constraints in~\eqref{eq:C1} and~\eqref{eq:C3}, they may not always maximize the rate.  
The above initial conditions yield an optimal scheme iff $t^{*}=D$, i.e., $g_t/f_t\leq g_D/f_D$ for all $t\in [D-1]$, achieving the rate $g_D/f_D$. 
Otherwise, if $g_t/f_t> g_D/f_D$ for some $t\in [D-1]$, our scheme achieves a rate greater than $g_D/f_D$.  

{\renewcommand{\arraystretch}{1.105}
\begin{table*}[!t]
    \centering
    \caption{The Query Table for the Case of $N=2$, $K=5$, and $D=2$\vspace{-0.125cm}}\label{tab:1}
    \scalebox{1.1}{
    \begin{tabular}{|c|c|c|}
    \hline
    $(i,j)$ & Server 1 & Server 2\\
    \hline
    \hline

    \multirow{1}{*}{$(1,0)$} & $a_{1}$, \dots, $a_{12}$, $b_1$, \dots, $b_{12}$ & $a_{42}$, \dots, $a_{53}$, $b_{42}$, \dots, $b_{53}$\\
    \hline
    
    \multirow{1}{*}{$(0,1)$} & $c_1$, \dots, $c_{12}$, $d_1$, \dots, $d_{12}$, $e_1$, \dots, $e_{12}$  & $c_{25}$, \dots $c_{36}$, $d_{25}$, \dots, $d_{36}$, $e_{25}$, \dots, $e_{36}$\\
    \hline
    \hline
    
    \multirow{1}{*}{$(2,0)$} & $a_{35}+b_{42}$, $a_{42}+b_{35}$, $a_{36}+b_{43}$, $a_{43}+b_{36}$, $a_{37}+b_{44}$ & $a_{76}+b_{1}$, $a_{1}+b_{76}$, $a_{77}+b_{2}$, $a_{2}+b_{77}$, $a_{78}+b_{3}$ \\
    \hline

    \multirow{6}{*}{$(1,1)$} & $a_{13}+c_{25}$, $a_{14}+c_{26}$, $a_{15}+c_{27}$, $a_{16}+c_{28}$, $a_{17}+c_{29}$ & $a_{54}+c_{1}$, $a_{55}+c_{2}$, $a_{56}+c_{3}$, $a_{57}+c_{4}$, $a_{58}+c_{5}$ \\
    & $a_{18}+d_{25}$, $a_{19}+d_{26}$, $a_{20}+d_{27}$, $a_{21}+d_{28}$, $a_{22}+d_{29}$ & $a_{59}+d_{1}$, $a_{60}+d_{2}$, $a_{61}+d_{3}$, $a_{62}+d_{4}$, $a_{63}+d_{5}$ \\
    & $a_{23}+e_{25}$, $a_{24}+e_{26}$, $a_{25}+e_{27}$, $a_{26}+e_{28}$, $a_{27}+e_{29}$ & $a_{64}+e_{1}$, $a_{65}+e_{2}$, $a_{66}+e_{3}$, $a_{67}+e_{4}$, $a_{68}+e_{5}$ \\
    & $b_{13}+c_{30}$, $b_{14}+c_{31}$, $b_{15}+c_{32}$, $b_{16}+c_{33}$, $b_{17}+c_{34}$ & $b_{54}+c_{6}$, $b_{55}+c_{7}$, $b_{56}+c_{8}$, $b_{57}+c_{9}$, $b_{58}+c_{10}$ \\
    & $b_{18}+d_{30}$, $b_{19}+d_{31}$, $b_{20}+d_{32}$, $b_{21}+d_{33}$, $b_{22}+d_{34}$ & $b_{59}+d_{6}$, $b_{60}+d_{7}$, $b_{61}+d_{8}$, $b_{62}+d_{9}$, $b_{63}+d_{10}$ \\
    & $b_{23}+e_{30}$, $b_{24}+e_{31}$, $b_{25}+e_{32}$, $b_{26}+e_{33}$, $b_{27}+e_{34}$ & $b_{64}+e_{6}$, $b_{65}+e_{7}$, $b_{66}+e_{8}$, $b_{67}+e_{9}$, $b_{68}+e_{10}$ \\
    \hline 

    \multirow{3}{*}{$(0,2)$} & $c_{13}+d_{13}$, $c_{14}+d_{14}$, $c_{15}+d_{15}$, $c_{16}+d_{16}$, $c_{17}+d_{17}$ & $c_{37}+d_{37}$, $c_{38}+d_{38}$, $c_{39}+d_{39}$, $c_{40}+d_{40}$, $c_{41}+d_{41}$ \\
    & $c_{18}+e_{13}$, $c_{19}+e_{14}$, $c_{20}+e_{15}$, $c_{21}+e_{16}$, $c_{22}+e_{17}$ & $c_{42}+e_{37}$, $c_{43}+e_{38}$, $c_{44}+e_{39}$, $c_{45}+e_{40}$, $c_{46}+e_{41}$ \\
    & $d_{18}+e_{18}$, $d_{19}+e_{19}$, $d_{20}+e_{20}$, $d_{21}+e_{21}$, $d_{22}+e_{22}$ & $d_{42}+e_{42}$, $d_{43}+e_{43}$, $d_{44}+e_{44}$, $d_{45}+e_{45}$, $d_{46}+e_{46}$ \\
    \hline
    \hline
    
    \multirow{3}{*}{$(2,1)$} & $a_{44}+b_{37}+c_{35}$, $a_{38}+b_{45}+c_{36}$ & $a_{3}+b_{78}+c_{11}$, $a_{79}+b_{4}+c_{12}$\\& $a_{45}+b_{38}+d_{35}$, $a_{39}+b_{46}+d_{36}$ & $a_{4}+b_{79}+d_{11}$, $a_{80}+b_{5}+d_{12}$\\& $a_{46}+b_{39}+e_{35}$, $a_{40}+b_{47}+e_{36}$ & $a_{5}+b_{80}+e_{11}$, $a_{81}+b_{6}+e_{12}$\\
    \hline
    
    \multirow{6}{*}{$(1,2)$} & $a_{28}+c_{37}+d_{37}$, $a_{29}+c_{38}+d_{38}$ & $a_{69}+c_{13}+d_{13}$, $a_{70}+c_{14}+d_{14}$\\
    & $a_{30}+c_{42}+e_{37}$, $a_{31}+c_{43}+e_{38}$ & $a_{71}+c_{18}+e_{13}$, $a_{72}+c_{19}+e_{14}$\\
    & $a_{32}+d_{42}+e_{42}$, $a_{33}+d_{43}+e_{43}$ & $a_{73}+d_{18}+e_{18}$, $a_{74}+d_{19}+e_{19}$\\
    & $b_{28}+c_{39}+d_{39}$, $b_{29}+c_{40}+d_{40}$ & $b_{69}+c_{15}+d_{15}$, $b_{70}+c_{16}+d_{16}$\\
    & $b_{30}+c_{44}+e_{39}$, $b_{31}+c_{45}+e_{40}$ & $b_{71}+c_{20}+e_{15}$, $b_{72}+c_{21}+e_{16}$\\
    & $b_{32}+d_{44}+e_{44}$, $b_{33}+d_{45}+e_{45}$ & $b_{73}+d_{20}+e_{20}$, $b_{74}+d_{21}+e_{21}$\\
    \hline

    \multirow{1}{*}{$(0,3)$} & $c_{23}+d_{23}+e_{23}$, $c_{24}+d_{24}+e_{24}$ & $c_{47}+d_{47}+e_{47}$, $c_{48}+d_{48}+e_{48}$\\
    \hline
    \hline

    \multirow{3}{*}{$(2,2)$} & $a_{47}+b_{40}+c_{41}+d_{41}$ & $a_{6}+b_{81}+c_{17}+d_{17}$\\
    & $a_{41}+b_{48}+c_{46}+e_{41}$ & $a_{82}+b_{7}+c_{22}+e_{17}$\\
    & $a_{48}+b_{41}+d_{46}+e_{46}$ & $a_{7}+b_{82}+d_{22}+e_{22}$\\
    \hline 
    
    \multirow{2}{*}{$(1,3)$} & $a_{34}+c_{47}+d_{47}+e_{47}$ & $a_{75}+c_{23}+d_{23}+e_{23}$\\ 
    & $b_{34}+c_{48}+d_{48}+e_{48}$ & $b_{75}+c_{24}+d_{24}+e_{24}$\\ 
    \hline
    
    \end{tabular}
    }\vspace{-0.125cm} 
\end{table*}
}

\section{An Illustrative Example}
In this section, we present an illustrative example of the proposed scheme for $N=2$, $K=5$, and $D=2$. 

For simplicity, 
we represent the demand messages by $a$ and $b$, and 
the interference messages by $c$, $d$, and $e$.

For this example, the vectors $\mathrm{v}_1, \dots,\mathrm{v}_5$ are given by $\mathrm{v}_5 = [0,1]$, $\mathrm{v}_4 = [1,0]$, $\mathrm{v}_3 = [2,1]$, $\mathrm{v}_2 = [5,2]$, and $\mathrm{v}_1 = [12,5]$,  
and the vectors $\mathrm{f}$ and $\mathrm{g}$ are given by 
$\mathrm{f} = [135,56]$ 
and 
$\mathrm{g} = [82,34]$. 
Since $g_1/f_1=82/135>g_2/f_2=34/56$, then $t^{*}=1$. 
Accordingly, $x_1=v_{1,t^{*}}/g_{t^{*}} = v_{1,1}/g_1 = 12/82$, 
$x_2=v_{2,t^{*}}/g_{t^{*}} = v_{2,1}/g_1 = 5/82$,
$x_3=v_{3,t^{*}}/g_{t^{*}} = v_{3,1}/g_1 = 2/82$,
$x_4 = 1/g_{t^{*}} = 1/g_1 = 1/82$, and $x_5 = 0$. 
Taking $L = S(N-1)^{K-D}g_{t^{*}} = 82$ for $S=1$, 
the scheme selects $L_1,\dots,L_5$ as follows: 
$L_1 = Lx_1 = 12$, $L_2 = Lx_2 = 5$, $L_3 = Lx_3 = 2$, $L_4 = Lx_4 = 1$, and $L_5 = Lx_5 = 0$. 
Note that, for these values of $L_1,\dots,L_5$, $R_1 = \sum_{j=0}^{3} \binom{3}{j} L_{1+j} = 34$ and $R_2 = \frac{1}{2}\sum_{j=0}^{3} \binom{3}{j}L_{2+j} = 7$, both of which are integers.

Table~\ref{tab:1} presents the queries constructed using the proposed scheme, where each message is randomly and independently divided into ${L=82}$ subpackets of equal size, labeled as $a_1,\dots,a_{82}$, $b_1,\dots,b_{82}$, $c_1,\dots,c_{82}$, 
$d_1,\dots,d_{82}$, and 
$e_1,\dots,e_{82}$. 
From the table, it can be observed that for each server, there are ${L_1 = 12}$ queries in the form of singletons for each message; 
${L_2 = 5}$ queries for each pair of messages, either in the form of $(2,0)$-sums, $(1,1)$-sums, or $(0,2)$-sums; 
${L_3 = 2}$ queries for each triple of messages, either in the form of $(2,1)$-sums, $(1,2)$-sums, or $(0,3)$-sums; 
and ${L_4 = 1}$ query for each quadruple of messages, either in the form of $(2,2)$-sums or $(1,3)$-sums. 
 
The privacy condition is satisfied since, from the perspective of each server, every message appears in an equal number of singletons, sums of two, sums of three, and sums of four, and there are no duplicate subpackets for any message in the answer from any server.

To prove recoverability, it suffices to show that the user can recover the subpackets $a_{1},\dots,a_{41}$ and $b_{1},\dots,b_{41}$ from Server~1 and $a_{42},\dots,a_{82}$ and $b_{42},\dots,b_{82}$ from Server~2. 
We will explain the recovery process for $a_{1},\dots,a_{41}$ and $b_{1},\dots,b_{41}$ from Server~1. 
The recovery process for $a_{42},\dots,a_{82}$ and $b_{42},\dots,b_{82}$ from Server~2 follows similarly. 

The subpackets $a_{1},\dots,a_{12}$ and $b_{1},\dots,b_{12}$ are retrieved directly from Server~1.
The subpackets $a_{13},\dots,a_{27}$ are recovered using the first half of the $(1,1)$-sums retrieved from Server~1, by canceling out  $c_{25},\dots,c_{29}$, $d_{25},\dots,d_{29}$, and $e_{25},\dots,e_{29}$ retrieved from Server~2, and 
the subpackets $b_{13},\dots,b_{27}$ are recovered using the second half of the $(1,1)$-sums retrieved from Server~1, by canceling out $c_{30},\dots,c_{34}$, $d_{30},\dots,d_{34}$, and $e_{30},\dots,e_{34}$ retrieved from Server~2.

The subpackets $a_{28},\dots,a_{33}$ are recovered using the first half of the $(1,2)$-sums retrieved from Server~1, by canceling out ${c_{37}+d_{37}}$, ${c_{38}+d_{38}}$, ${c_{42}+e_{37}}$, ${c_{43}+e_{38}}$, ${d_{42}+e_{42}}$, and ${d_{43}+e_{43}}$ retrieved from Server~2, and 
the subpackets $b_{28},\dots,b_{33}$ are recovered using the second half of the $(1,2)$-sums retrieved from Server~1, by canceling out ${c_{39}+d_{39}}$, ${c_{40}+d_{40}}$, ${c_{44}+e_{39}}$, ${c_{45}+e_{40}}$, ${d_{44}+e_{44}}$, and ${d_{45}+e_{45}}$ retrieved from Server~2. 

The subpackets $a_{34}$ and $b_{34}$ are recovered using the $(1,3)$-sums retrieved from Server~1, by canceling out $c_{47}+d_{47}+e_{47}$ and $c_{48}+d_{48}+e_{48}$ retrieved from Server~2.

The subpackets $a_{35}$, $a_{36}$, and $a_{37}$ are recovered using three of the $(2,0)$-sums retrieved from Server~1, by canceling out $b_{42}$, $b_{43}$, and $b_{44}$ retrieved from Server~2, and 
the subpackets $b_{35}$ and $b_{36}$ are recovered using the other two $(2,0)$-sums retrieved from Server~1, by canceling out $a_{42}$ and $a_{43}$ retrieved from Server~2.

The subpackets $a_{38}$, $a_{39}$, and $a_{40}$ are recovered using three of the $(2,1)$-sums retrieved from Server~1, by canceling out $b_{45},b_{46},b_{47}$ and $c_{36},d_{36},e_{36}$ retrieved from Server~2, and 
the subpackets $b_{37}$, $b_{38}$, and $b_{39}$ are recovered using the other three $(2,1)$-sums retrieved from Server~1, by canceling out $a_{44},a_{45},a_{46}$ and $c_{35},d_{35},e_{35}$ retrieved from Server~2. 

The subpacket $a_{41}$ is recovered using one of the $(2,2)$-sums retrieved from Server~1, by canceling out $b_{48}$ and ${c_{46}+e_{41}}$ retrieved from Server~2, and 
the subpackets $b_{40}$ and $b_{41}$ are recovered using the other two $(2,2)$-sums retrieved from Server~1, by canceling out $a_{47}$, $a_{48}$,   ${c_{41}+d_{41}}$, and ${d_{46}+e_{46}}$ retrieved from Server~2.

For this example, the user downloads a total of ${K\times L_1 = 60}$ singletons, ${\binom{K}{2} L_2 = 50}$ sums of two, ${\binom{K}{3} L_3 = 20}$ sums of three, and ${\binom{K}{4} L_4 = 5}$ sums of four from each server, where each download has the same size as a message subpacket. 
This yields downloading ${M = 135}$ subpackets from each server. 
Since there are ${D=2}$ demand messages, each containing ${L=82}$ subpackets, 
the rate of the scheme is $\frac{DL}{NM} = \frac{82}{135}\approx 0.6074$. 
For the same instance, the scheme in~\cite{BU2018} achieves a lower rate of $\frac{17}{28}\approx 0.6071$  (for details, see \cite[Section~5.1]{BU2018}).

\bibliographystyle{IEEEtran}
\bibliography{PIR_PC_Refs}

\begin{thebibliography}{10}
\providecommand{\url}[1]{#1}
\csname url@samestyle\endcsname
\providecommand{\newblock}{\relax}
\providecommand{\bibinfo}[2]{#2}
\providecommand{\BIBentrySTDinterwordspacing}{\spaceskip=0pt\relax}
\providecommand{\BIBentryALTinterwordstretchfactor}{4}
\providecommand{\BIBentryALTinterwordspacing}{\spaceskip=\fontdimen2\font plus
\BIBentryALTinterwordstretchfactor\fontdimen3\font minus \fontdimen4\font\relax}
\providecommand{\BIBforeignlanguage}[2]{{%
\expandafter\ifx\csname l@#1\endcsname\relax
\typeout{** WARNING: IEEEtran.bst: No hyphenation pattern has been}%
\typeout{** loaded for the language `#1'. Using the pattern for}%
\typeout{** the default language instead.}%
\else
\language=\csname l@#1\endcsname
\fi
#2}}
\providecommand{\BIBdecl}{\relax}
\BIBdecl

\bibitem{SJ2017}
H.~Sun and S.~A. Jafar, ``{The Capacity of Private Information Retrieval},'' \emph{IEEE Transactions on Information Theory}, vol.~63, no.~7, pp. 4075--4088, July 2017.

\bibitem{SJ2016ArbitraryTIFS}
------, ``Optimal download cost of private information retrieval for arbitrary message length,'' \emph{IEEE Transactions on Information Forensics and Security}, vol.~12, no.~12, pp. 2920--2932, 2017.

\bibitem{SJ2018Multiround}
------, ``{Multiround Private Information Retrieval: Capacity and Storage Overhead},'' \emph{IEEE Transactions on Information Theory}, vol.~64, no.~8, pp. 5743--5754, 2018.

\bibitem{ZX2019}
Z.~Zhang and J.~Xu, ``{The Optimal Sub-Packetization of Linear Capacity-Achieving PIR Schemes with Colluding Servers},'' \emph{IEEE Transactions on Information Theory}, vol.~65, no.~5, pp. 2723--2735, 2019.

\bibitem{TSC2019}
C.~Tian, H.~Sun, and J.~Chen, ``{Capacity-Achieving Private Information Retrieval Codes with Optimal Message Size and Upload Cost},'' \emph{IEEE Transactions on Information Theory}, vol.~65, no.~11, pp. 7613--7627, 2019.

\bibitem{BU2018}
K.~{Banawan} and S.~{Ulukus}, ``{Multi-Message Private Information Retrieval: Capacity Results and Near-Optimal Schemes},'' \emph{IEEE Transactions on Information Theory}, vol.~64, no.~10, pp. 6842--6862, Oct 2018.

\bibitem{WHS2022}
N.~Wang, A.~Heidarzadeh, and A.~Sprintson, ``{Multi-Message Private Information Retrieval: A Scalar Linear Solution},'' in \emph{58th Annual Allerton Conference on Communication, Control, and Computing}, 2022, pp. 1--8.

\bibitem{VBU2022}
S.~Vithana, K.~Banawan, and S.~Ulukus, ``{Semantic Private Information Retrieval},'' \emph{IEEE Transactions on Information Theory}, vol.~68, no.~4, pp. 2635--2652, 2022.

\bibitem{SJ2018Colluding}
H.~Sun and S.~A. Jafar, ``{The Capacity of Robust Private Information Retrieval with Colluding Databases},'' \emph{IEEE Transactions on Information Theory}, vol.~64, no.~4, pp. 2361--2370, 2018.

\bibitem{BU2019Colluding}
K.~Banawan and S.~Ulukus, ``{The Capacity of Private Information Retrieval from Byzantine and Colluding Databases},'' \emph{IEEE Transactions on Information Theory}, vol.~65, no.~2, pp. 1206--1219, 2019.

\bibitem{LJJ2021}
Y.~Lu, Z.~Jia, and S.~A. Jafar, ``{Double Blind T-Private Information Retrieval},'' \emph{IEEE Journal on Selected Areas in Information Theory}, vol.~2, no.~1, pp. 428--440, 2021.

\bibitem{HFLH2022}
L.~Holzbaur, R.~Freij-Hollanti, J.~Li, and C.~Hollanti, ``{Toward the Capacity of Private Information Retrieval From Coded and Colluding Servers},'' \emph{IEEE Transactions on Information Theory}, vol.~68, no.~1, pp. 517--537, 2022.

\bibitem{TGKHHER2017}
R.~Tajeddine, O.~W. Gnilke, D.~Karpuk, R.~Freij-Hollanti, C.~Hollanti, and S.~E. Rouayheb, ``{Private Information Retrieval Schemes for Coded Data with Arbitrary Collusion Patterns},'' in \emph{IEEE International Symposium on Information Theory}, June 2017, pp. 1908--1912.

\bibitem{BU18}
K.~Banawan and S.~Ulukus, ``{The Capacity of Private Information Retrieval from Coded Databases},'' \emph{IEEE Transactions on Information Theory}, vol.~64, no.~3, pp. 1945--1956, March 2018.

\bibitem{BAWU2020}
K.~Banawan, B.~Arasli, Y.-P. Wei, and S.~Ulukus, ``{The Capacity of Private Information Retrieval From Heterogeneous Uncoded Caching Databases},'' \emph{IEEE Transactions on Information Theory}, vol.~66, no.~6, pp. 3407--3416, 2020.

\bibitem{SSM2018}
S.~P. {Shariatpanahi}, M.~J. {Siavoshani}, and M.~A. {Maddah-Ali}, ``{Multi-Message Private Information Retrieval with Private Side Information},'' in \emph{IEEE Information Theory Workshop}, 2018.

\bibitem{MMM2019}
M.~H. Mousavi, M.~Ali Maddah-Ali, and M.~Mirmohseni, ``Private inner product retrieval for distributed machine learning,'' in \emph{IEEE International Symposium on Information Theory}, 2019, pp. 355--359.

\end{thebibliography}

\appendix
In this appendix, we show that when $D=2$, our scheme outperforms the scheme in~\cite{BU2018} for all odd $K$, and performs similarly for all even $K$. 
Specifically, we show that $g_1/f_1 > g_2/f_2$ for all odd $K$, and $g_1/f_1 = g_2/f_2$ for all even $K$. 

To simplify the notation, we denote by $\alpha_s$ and $\beta_s$ the two coordinates of the vector $\mathrm{v}_s$ for each ${s\in [K]}$, 
i.e., $\mathrm{v}_s = [\alpha_s,\beta_s]$. 
Note that 
${\alpha_{K-1} = 1}$ and ${\beta_{K-1} = 0}$ since ${\mathrm{v}_{K-1} = [1,0]}$, 
and 
${\alpha_{K} = 0}$ and ${\beta_{K} = 1}$ since ${\mathrm{v}_{K} = [0,1]}$. 

Using~\eqref{eq:f} and~\eqref{eq:g}, we can write $g_1/f_1$ and $g_2/f_2$ as follows: 
\begin{equation}\label{eq:g1f1}
\frac{g_1}{f_1} = 1- \frac{\sum_{s=1}^{K-2} \binom{K-2}{s}\alpha_s}{\sum_{s=1}^{K} \binom{K}{s}\alpha_s},
\end{equation}
and 
\begin{equation}\label{eq:g2f2}
\frac{g_2}{f_2} = 1- \frac{\sum_{s=1}^{K-2} \binom{K-2}{s}\beta_s}{\sum_{s=1}^{K} \binom{K}{s}\beta_s}. 
\end{equation}

First, we show that 
\[\sum_{s=1}^{K-2}\binom{K-2}{s} \alpha_s = \frac{1}{N}\sum_{i=1}^{K}\binom{K}{i}\alpha_i - \frac{1}{N}\sum_{t=1}^{2} \binom{2}{t}\alpha_t.\] 

From the definition of $\mathrm{v}_s$, we know that
\[\alpha_s = \frac{1}{N-1}\sum_{t=1}^{2} \binom{2}{t} \alpha_{t+s}\] for each $s\in [K-2]$. 
Thus, 
\begin{align*}
&\sum_{s=1}^{K-2}\binom{K-2}{s} \alpha_s = \frac{1}{N-1}\sum_{s=1}^{K-2}\sum_{t=1}^{2}\binom{K-2}{s}\binom{2}{t} \alpha_{t+s}.
\end{align*}
Also, we know that
\[\sum_{s=0}^{K-2}\sum_{t=0}^{2}\binom{K-2}{s}\binom{2}{t} \alpha_{t+s} = \sum_{i=0}^{K} \binom{K}{i} \alpha_i,\] 
where $\alpha_0 := 0$. 
Subtracting the terms for $s=0$ or $t=0$, 
\begin{align*}
& \sum_{s=1}^{K-2}\sum_{t=1}^{2}\binom{K-2}{s}\binom{2}{t} \alpha_{t+s}\\ 
& \quad = \sum_{i=1}^{K} \binom{K}{i} \alpha_i - \sum_{t=0}^{2}\binom{2}{t}\alpha_t - \sum_{s=1}^{K-2}\binom{K-2}{s}\alpha_s.
\end{align*}
By these equations, we have
\begin{dmath*}
\sum_{s=1}^{K-2}\binom{K-2}{s} \alpha_s = \frac{1}{N-1}\sum_{s=1}^{K} \binom{K}{s} \alpha_s - \frac{1}{N-1}\sum_{t=1}^{2} \binom{2}{t} \alpha_t - \frac{1}{N-1}\sum_{s=1}^{K-2} \binom{K-2}{s} \alpha_s, 
\end{dmath*} which can be rewritten as
\begin{align*}
\sum_{s=1}^{K-2}\binom{K-2}{s} \alpha_s & = \frac{1}{N}\sum_{i=1}^{K} \binom{K}{i} \alpha_i - \frac{1}{N}\sum_{t=1}^{2} \binom{2}{t} \alpha_t.
\end{align*} 

Similarly, it can be shown that 
\[\sum_{s=1}^{K-2}\binom{K-2}{s} \beta_s = \frac{1}{N}\sum_{i=1}^{K}\binom{K}{i}\beta_i - \frac{1}{N}\sum_{t=1}^{2}\binom{2}{t}\beta_t.\] 

Substituting these into~\eqref{eq:g1f1} and~\eqref{eq:g2f2} and rearranging the terms,  
it remains to show that
\begin{dmath}\label{eq:Ineq1}
\sum_{i=1}^{K}\binom{K}{i} \left(\left(\alpha_1\beta_i-\alpha_i\beta_1\right)-\frac{1}{2}\left(\alpha_i\beta_2-\alpha_2\beta_i\right)\right)>0
\end{dmath}
for all odd $K$, 
and 
\begin{dmath}\label{eq:Ineq2}
\sum_{i=1}^{K}\binom{K}{i} \left(\left(\alpha_1\beta_i-\alpha_i\beta_1\right)-\frac{1}{2}\left(\alpha_i\beta_2-\alpha_2\beta_i\right)\right)=0
\end{dmath}
for all even $K$. 

To prove these results, we rely on the following two identities,  which hold for all $i\in [K]$: 
\begin{equation}\label{eq:ab1}
\alpha_1\beta_i-\alpha_i\beta_1 = \left(\frac{1}{1-N}\right)^{K-i}\alpha_{K-i+1},
\end{equation}
and 
\begin{equation}\label{eq:ab2}
\alpha_i\beta_2-\alpha_2\beta_i = \left(\frac{1}{1-N}\right)^{K-i-1}\beta_{K-i+1}. 
\end{equation}

To verify these identities, we use the relation
\begin{equation}\label{eq:rel}
\alpha_s = \frac{2}{N
-1}\alpha_{s+1}+\frac{1}{N-1}\alpha_{s+2}
\end{equation} 
for each $s\in [K-2]$. 
Since $\alpha_{K-1} = 1$ and $\alpha_{K}=0$, 
the closed form expression for $\alpha_s$ is given by  
\begin{equation}\label{eq:alphas}
\alpha_s = c_1 r_1^{K-s+1}+c_2 r_2^{K-s+1}, \quad \forall s\in [K],
\end{equation}
where \[{r_1:=\frac{1+\sqrt{N}}{N-1}}\] and \[{r_2:=\frac{1-\sqrt{N}}{N-1}}\] are the roots of the characteristic equation \[{r^2-\frac{2}{N-1}r-\frac{1}{N-1} = 0},\] 
and the constants \[{c_1:= \frac{r_2}{r_2-r_1}(N-1)}\] and \[{c_2:= \frac{r_1}{r_1-r_2}(N-1)}\] are determined by the conditions 
$\alpha_{K-1} = 1$ and $\alpha_{K} = 0$, 
i.e., ${c_1 r_1^2+c_2 r_2^2 = 1}$ and $c_1 r_1+c_2 r_2 = 0$.
Similarly, the closed form expression for $\beta_s$ is given by  
\begin{equation}\label{eq:betas}
\beta_s = c'_1 r_1^{K-s+1}+c'_2 r_2^{K-s+1}, \quad \forall s\in [K],
\end{equation}
where ${r_1}$ and ${r_2}$ are defined as before, 
and the constants \[{c'_1:= \frac{r_2^2}{r_1-r_2}(N-1)}\] and \[{c'_2:= \frac{r_1^2}{r_2-r_1}(N-1)}\] are determined by the conditions 
$\alpha_{K-1} = 0$ and $\alpha_{K} = 1$, 
i.e., ${c'_1 r_1^2+c'_2 r_2^2 = 0}$ and $c'_1 r_1+c'_2 r_2 = 1$.

Using~\eqref{eq:alphas}, we can write 
\begin{align*}
\alpha_1\beta_i - \alpha_i\beta_1 & = (c_1r_1^{K}+c_2r_2^{K})(c'_1r_1^{K-i+1}+c'_2r_2^{K-i+1})\\
& \quad - (c_1r_1^{K-i+1}+c_2r_2^{K-i+1})(c'_1r_1^{K}+c'_2r_2^{K})\\
& = c_1c'_2 r_1^K r_2^{K-i+1} + c'_1c_2r_1^{K-i+1}r_2^K\\ 
& \quad - c_1c'_2r_1^{K-i+1}r_2^K - c'_1c_2 r_1^K r_2^{K-i+1}\\
& = c_1c'_2 r_1^{K-i+1}r_2^{K-i+1} (r_1^{i-1}-r_2^{i-1})\\
& \quad - c'_1c_2 r_1^{K-i+1}r_2^{K-i+1} (r_1^{i-1}-r_2^{i-1}) \\ 
& = (r_1r_2)^{K-i+1}(c_1c'_2-c'_1c_2)(r_1^{i-1}-r_2^{i-1}).
\end{align*}
Since
\[r_1 r_2 = \frac{1}{1-N}\] and \[c_1c'_2 - c'_1c_2 = \frac{1-N}{r_1-r_2},\] 
it follows that
\[\alpha_1\beta_i - \alpha_i\beta_1 = \left(\frac{1}{1-N}\right)^{K-i}\left(\frac{r_1^{i-1}-r_2^{i-1}}{r_1-r_2}\right).\]
Moreover, 
\[\alpha_{K-i+1} = c_1 r_1^{i} + c_2 r_2^{i} = \frac{r_1^{i-1}-r_2^{i-1}}{r_1-r_2}.\] 
Thus, 
\[\alpha_1\beta_i - \alpha_i\beta_1 = \left(\frac{1}{1-N}\right)^{K-i} \alpha_{K-i+1}.\] 

Similarly, 
using~\eqref{eq:betas}, 
we can write 
\begin{align*}
\alpha_i\beta_2 - \alpha_2\beta_i & = (c_1r_1^{K-i+1}+c_2r_2^{K-i+1})(c'_1r_1^{K-1}+c'_2r_2^{K-1})\\
& \quad - (c_1r_1^{K-1}+c_2r_2^{K-1})(c'_1r_1^{K-i+1}+c'_2r_2^{K-i+1})\\
& = c_1c'_2 r_1^{K-i+1} r_2^{K-1} + c'_1c_2r_1^{K-1}r_2^{K-i+1}\\ 
& \quad - c_1c'_2r_1^{K-1}r_2^{K-i+1} - c'_1c_2 r_1^{K-i+1} r_2^{K-1}\\
& = c_1c'_2 r_1^{K-i}r_2^{K-i} (r_1r_2^{i-1}-r_1^{i-1}r_2)\\
& \quad - c'_1c_2 r_1^{K-i}r_2^{K-i} (r_1 r_2^{i-1}-r_1^{i-1}r_2) \\ 
& = (r_1r_2)^{K-i}(c_1c'_2-c'_1c_2)(r_1r_2^{i-1}-r_1^{i-1}r_2).
\end{align*}
Similarly, 
it follows that
\[\alpha_i\beta_2 - \alpha_2\beta_i = \left(\frac{1}{1-N}\right)^{K-i-1}\left(\frac{r_1r_2^{i-1}-r_1^{i-1}r_2}{r_1-r_2}\right).\]
Moreover, 
\[\beta_{K-i+1} = c'_1 r_1^{i} + c'_2 r_2^{i} = \frac{r_1r_2^{i-1}-r_1^{i-1}r_2}{r_1-r_2}.\] 
Thus, 
\[\alpha_i\beta_2 - \alpha_2\beta_i = \left(\frac{1}{1-N}\right)^{K-i-1} \beta_{K-i+1}.\]

Using~\eqref{eq:ab1} and~\eqref{eq:ab2}, 
we can write
\begin{align}\label{eq:BigRel}
& \sum_{i=1}^{K}\binom{K}{i} \left(\left(\alpha_1\beta_i-\alpha_i\beta_1\right)-\frac{1}{2}\left(\alpha_i\beta_2-\alpha_2\beta_i\right)\right) \nonumber \\
& \quad = \sum_{i=1}^{K} \binom{K}{i} \left(\frac{1}{1-N}\right)^{K-i}\left(\alpha_{K-i+1} +\frac{N-1}{2}\beta_{K-i+1}\right) \nonumber \\ 
& \quad = \sum_{s=1}^{K} \binom{K}{s-1} \left(\frac{1}{1-N}\right)^{s-1}\left(\alpha_{s} +\frac{N-1}{2}\beta_{s}\right).
\end{align}

To simplify this further, we use the following identity, which holds for all $s\in [K-1]$:  
\[\beta_{s} = \frac{1}{N-1}\alpha_{s+1}.\]
This identity holds because 
\begin{align*}
\beta_{s} & = c'_1 r_1^{K-s+1}+c'_2 r_2^{K-s+1}\\
& = (N-1) \left(\frac{r_1^{K-s+1} r_2^2-r_1^2 r_2^{K-s+1}}{r_1-r_2}\right)\\
& = (N-1)(r_1r_2)^2 \left(\frac{r_1^{K-s-1}-r_2^{K-s-1}}{r_1-r_2}\right)\\
& = \frac{1}{N-1}\left(\frac{r_1^{K-s-1}-r_2^{K-s-1}}{r_1-r_2}\right)
\end{align*} 
and 
\[\alpha_{s+1} = c_1 r_1^{K-s}+c_2 r_2^{K-s} = \frac{r_1^{K-s-1}-r_2^{K-s-1}}{r_1-r_2}\]
for each $s\in [K-1]$. 

Rewriting~\eqref{eq:BigRel} by using this identity, 
we have 
\begin{align*}
& \sum_{i=1}^{K}\binom{K}{i} \left(\left(\alpha_1\beta_i-\alpha_i\beta_1\right)-\frac{1}{2}\left(\alpha_i\beta_2-\alpha_2\beta_i\right)\right) \\
& \quad = \sum_{s=1}^{K-1} \binom{K}{s-1} \left(\frac{1}{1-N}\right)^{s-1}\left(\alpha_{s} +\frac{1}{2}\alpha_{s+1}\right) \\ 
& \quad \quad + K\left(\frac{1}{1-N}\right)^{K-1}\left(\alpha_K+\frac{N-1}{2}\beta_K\right).  
\end{align*}
Since   
\[\alpha_{s}+\frac{1}{2}\alpha_{s+1} = \frac{N-1}{2}\alpha_{s-1}\] for all $s\in [2:K-1]$, and $\alpha_K=0$, and $\beta_K = \alpha_{K-1}$,  
we can write  
\begin{align*}
& \sum_{i=1}^{K}\binom{K}{i} \left(\left(\alpha_1\beta_i-\alpha_i\beta_1\right)-\frac{1}{2}\left(\alpha_i\beta_2-\alpha_2\beta_i\right)\right) \\
& \quad = \alpha_1+\frac{1}{2}\alpha_2 + \frac{N-1}{2}\sum_{s=1}^{K} \binom{K}{s} \left(\frac{1}{1-N}\right)^{s}\alpha_{s}.  
\end{align*}

Since \[\alpha_s = \frac{r_1^{K-s}-r_2^{K-s}}{r_1-r_2}\] for all $s\in [K]$, and 
$r_1$ and $r_2$ are the roots of the equation \[r^2 - \frac{2}{N-1}r - \frac{1}{N-1} = 0,\] or equivalently, \[\frac{N-1}{2}r^2 - r - \frac{1}{2} = 0,\] 
we have
\begin{align*}
\alpha_1+\frac{1}{2}\alpha_2 & = \frac{\left(r_1^{K-1}-r_2^{K-1}\right) + \frac{1}{2}\left(r_1^{K-2}-r_2^{K-2}\right)}{r_1-r_2} \\
& = \frac{r_1^{K-2}\left(r_1+\frac{1}{2}\right) - r_2^{K-2}\left(r_2+\frac{1}{2}\right)}{r_1-r_2} \\
& = \frac{N-1}{2}\left(\frac{r_1^K-r_2^K}{r_1-r_2}\right).
\end{align*}
Similarly, we have
\begin{align*}
& \frac{N-1}{2}\sum_{s=1}^{K} \binom{K}{s} \left(\frac{1}{1-N}\right)^{s}\alpha_{s} \\
& \quad = \frac{N-1}{2}\sum_{s=1}^{K} \binom{K}{s} \left(\frac{1}{1-N}\right)^{s} \left(\frac{r_1^{K-s}-r_2^{K-s}}{r_1-r_2}\right). 
\end{align*}
Using the binomial theorem, 
\[\sum_{s=1}^{K}\binom{K}{s} \left(\frac{1}{1-N}\right)^{s} r_1^{K-s} = \left(r_1+\frac{1}{1-N}\right)^{K} - r_1^K,\] 
and 
\[\sum_{s=1}^{K}\binom{K}{s} \left(\frac{1}{1-N}\right)^{s} r_2^{K-s} = \left(r_2+\frac{1}{1-N}\right)^{K} - r_2^K.\]
Thus, we have
\begin{align*}
& \sum_{i=1}^{K}\binom{K}{i} \left(\left(\alpha_1\beta_i-\alpha_i\beta_1\right)-\frac{1}{2}\left(\alpha_i\beta_2-\alpha_2\beta_i\right)\right) \\
& \quad = \frac{N-1}{2}\left(\frac{r_1^K-r_2^K}{r_1-r_2}\right) \\
& \quad \quad  + \frac{N-1}{2}\left(\frac{1}{r_1-r_2}\right) \left(\left(r_1+\frac{1}{1-N}\right)^{K} -r_1^K\right.  \\ 
& \quad \quad \quad \quad \quad \quad \quad \quad \quad \quad \quad \quad \left. -\left(r_2+\frac{1}{1-N}\right)^{K}+ r_2^K \right) \\
& \quad = \frac{N-1}{2} \left(\frac{1}{r_1-r_2}\right)\\
& \quad \quad \times \left(\left(r_1+\frac{1}{1-N}\right)^{K} - \left(r_2+\frac{1}{1-N}\right)^{K}\right).    
\end{align*}

Since
\[\frac{N-1}{2}\left(\frac{1}{r_1-r_2}\right) = \frac{(N-1)^2}{4\sqrt{N}}>0,\]
it suffices to show that 
\[\left(r_1+\frac{1}{1-N}\right)^{K} - \left(r_2+\frac{1}{1-N}\right)^{K} >0\]
for all odd $K$, and 
\[\left(r_1+\frac{1}{1-N}\right)^{K} - \left(r_2+\frac{1}{1-N}\right)^{K} = 0\] for all even $K$. 
This is immediate because 
\begin{align*}
& \left(r_1+\frac{1}{1-N}\right)^{K} - \left(r_2+\frac{1}{1-N}\right)^{K} \\
& \quad = \left(\frac{\sqrt{N}}{N-1}\right)^K  \left( 1- (-1)^K\right), 
\end{align*} 
where ${(1-(-1)^K) = 2>0}$ for odd $K$, and ${(1-(-1)^K) = 0}$ for even $K$. 
This completes the proof. 

\end{document}